\documentstyle[psfig,12pt,aaspp4,epsf]{article}

\def\arcmin{\hbox{$^\prime$}}
\def\arcsec{\hbox{$^{\prime\prime}$}}
\def\fmag{\hbox{$.\!\!^m$}}
\def\farcmin{\hbox{$.\!\!^\prime$}}

\begin{document}

\lefthead{Tovmassian et al.}
\righthead{Radio emission of ShCGs}

\slugcomment{{\it The Astrophysical Journal Supplement Series} in press}

\title{Radio emission of Shakhbazian Compact Galaxy Groups}

\author{H.M. Tovmassian, and V.H. Chavushyan}

\affil{Instituto Nacional de Astrof\'{\i}sica \'Optica y Electr\'onica, Apartado Postal 51 y 216, \\ 
Puebla, Pue, ZP 72000, M\'exico \\
Electronic mail: hrant@inaoep.mx and vahram@inaoep.mx}

\author{O.V. Verkhodanov}

\affil{Special Astrophysical Observatory RAS, Nizhnij Arkhyz, Karachai-Cherkessia, 351147, Russia \\
Electronic mail: vo@sao.ru}

\and

\author{H. Tiersch}

\affil{Sternwarte K\"{o}nigsleiten, M\"{u}nchen, Germany \\
Electronic mail: sternwarte.kgl@aon.at}

\begin{abstract}

Three hundred fifty three radio sources from the NRAO VLA Sky Survey (NVSS)  (Condon et al. \cite{con98}) and the FIRST Survey (White et al. \cite{white}), are detected in the areas of 179 Shakhbazian Compact Groups (ShCGs) of galaxies. Ninety three of them are identified with galaxies in 74 ShCGs. Six radio sources have complex structure. The radio spectra of 22 sources are determined. Radio luminosities of galaxies in ShCGs are in general higher than that of galaxies in Hickson Compact Groups (HCGs). 

The comparison of radio (at 1.4 GHz) and FIR (at 60 $\mu$m) fluxes of ShCG galaxies with that of HCG galaxies shows that galaxies in ShCGs are relatively stronger emitters at radio wavelengths, while galaxies in HCGs have relatively stronger FIR emission. The reasons of such difference is discussed.

\end{abstract}


\section{Introduction}

A couple of decades ago it was found that radio emission is observed by about three times more often from double galaxies and members of groups of galaxies than from single, isolated galaxies (Tovmassian \cite{tov69}; Sulentic \cite{sul76}). It is widely accepted that galactic nuclear activity, one of the manifestations of which is a radio emission, is a result of interaction between galaxies. Events of interaction should take place much more often in dense environments of Compact Groups (CGs) of galaxies. Therefore it was expected that some of member galaxies in CGs should be relatively strong radio emitters. The first systematic search for radio emission from Hickson Compact Groups (HCGs), revealed a radio emission at 18 cm above a flux limit of 1.5 mJy from 41 out of 88 observed member galaxies (Menon \& Hickson \cite{menh85}). It was found (Tovmassian \& Shahbazian \cite{tovsh81}) that in groups of galaxies the radio emission is most often observed from the optically first-ranked galaxies. Menon (\cite{men92}) showed that the radio emission in HCGs is almost always associated with the first-ranked {\it elliptical} galaxies, while the radio detected {\it spirals} are uniformly distributed among the three brightest members of a group.

Prior to HCGs the lists of the so called Shakhbasian Compact Groups of Compact Galaxies (SCGCGs) were published (Shakhbazian \cite{shakh73}, Baier \& Tiersch \cite{baiti79}, and references therein). It is the largest and a relatively homogeneous sample of galaxy groups known up to now. Shakhbazian's CGs were found on the POSS prints by an eye search. They were identified according to the following criteria:

- the groups consist of 5 - 15 members;\par 
- the apparent magnitude of the individual galaxies is between $14^m$ and $19^m$;\par
- the groups are compact, i.e. the distances between galaxies are only 3 - 5 times the diameters of galaxies;\par
- nearly all member galaxies are extremly red, at most 1-2 blue galaxies in a group is an exception;\par
- members of groups are compact (relatively high surface brightness, borders not diffuse);\par
- the groups are isolated.

ShCGs originally were called compact groups of {\it compact} galaxies, because the images of most of the constituent galaxies in these groups seemed very compact. Later observations of these groups with high angular resolution revealed that member galaxies mostly are of E and S0 types. The fraction of early type galaxies is 77\% in SCGs compared to 51\% in HCGs and 40\% in field galaxies (Tiersch et al. \cite{ti96a}).  Member galaxies in these groups are significantly redder than the field galaxies of the same morphological type. The difference in B-V and V-R is by $\sim 0.2\fmag$ redder compared to galaxies of the RC3. Anyhow, there were found emission-line, and even Seyfert galaxies among Shakhbazian groups members. 

Since, as it was found out, members of Shakhbazian groups turned to be not "compact", these groups in recent papers were called Shakhbazian Compact Groups of Galaxies (SCGGs) (Tovmassian et al. \cite{tov98a}, hereafter TMTST, \cite{tov98b}), or in analogy with HCGs, - Shakhbazian Compact Groups (ShCGs) (Tovmassian et al. \cite{tov98c}). Galaxies in ShCGs are relatively weak, because they are at least by three times further than HCGs (TMTST). Mainly for this reason, and probably for very unusual supposed composition the ShCGs attracted little attention until recently. The redshifts of only about 70 relatively bright and nearby ShCGs have been measured so far (Robinson \& Wampler \cite{row}; Arp et al. \cite{arp73}; Mirzoyan et al. \cite{mirz}; Amirkhanian \& Egikian \cite{amir87}; Amirkhanian \cite{amir89}; Kodaira et al. \cite{kod88}, \cite{kod90}; Kodaira \& Sekiguchi \cite{kods91}; Lynds et al. \cite{lynd}; del Olmo \& Moles \cite{delolm}; Tiersch et al. \cite{ti97}).

The densities of galaxies in ShCGs approaches to $10^{3}-10^{4}$ Mpc$^{-3}$. Hence the interaction and merging processes should be frequent in them. As a consequence some ShCGs could be a radio and FIR emitters. However due to larger distances and different morphological content (Tiersh et al. \cite{ti96a}) we could expect smaller rate of radio and FIR detections from ShCGs in comparison with HCGs. The results of the search for FIR emission of ShCGs were published recently by TMTST. In this paper we present the results of a search for radio emission from ShCGs.

\section{Observational data}

Three hundred seventy seven groups were included in the original lists of ShCGs. The positions of the group centers were given with accuracies of about $1\arcmin$. For identification of radio sources with ShCGs we used accurate coordinates of member galaxies measured afterwards from the Digitized Sky Survey (DSS) by Stoll et al. (\cite{st97}, and references therein). Positions of four groups, ShCG 206, 241, 301 and 353, were largely incorrect in original lists, and they were not found during accurate position measurements. It was revealed also that the group ShCG 214 coincides with ShCG 252, and the group ShCG 340 consists of two groups, ShCG 340a and ShCG 340b. Spectral observations showed that five groups, ShCG 12, 13, 78, 146, and 180, consist mainly of stars. Thus we searched for a radio emission at the positions of 367 ShCGs.

For looking for a radio emission from ShCGs we first examined the NRAO VLA Sky Survey (NVSS) (Condon et al. \cite{con98}) at 1.4 GHz at the positions of all 367 ShCGs. Generally an area $\pm3^{'}$ centered on each group was investigated. The FIRST Survey (White et al. \cite{white}), made at the same frequency, was used in addition for those 175 ShCGs that are located in the area of the sky, covered by the Survey until 1st November 1998.  

The errors of the radio position measurements in the used Surveys are small enough. For the NVSS sources they vary from $\leq1\arcsec$ for sources stronger than 15 mJy, to $7\arcsec$ at the Survey flux limit at $F\sim2.5$ mJy (Condon et al. \cite{con98}). For the FIRST radio sources the errors are smaller: $\leq0.5\arcsec$ for sources with flux densities $>3$ mJy, and $1\arcsec$ at the survey treshold of 1 mJy (White et al. \cite{white}).

\section{RESULTS AND DISCUSSION}

\subsection{Identifications}

Three hundred fifty three NVSS radio sources were found within boundaries of 179 ShCGs. Fifty three of these sources were registered also in the FIRST. Seven more sources were found only in the FIRST. The more correct positions and fluxes on the latter radio sources from the FIRST were used in the farther discussion. 

High positional accuracies of the used radio surveys allowed to identify with high confidence some of the found radio sources with certain galaxies in dense environments of ShCGs. The positions of these radio sources coincide appreciably well, generally within $2\arcsec-3\arcsec$, with corresponding objects in ShCGs.

The probability of the reality of radio identifications was estimated by the "Likelihood Ratio" (LR) (de Ruiter et al. \cite{ruit}). It was assumed that the density of background radio sources is $\rho=5.16 \times 10^{-4}$ per $arcsec^2$ for high galactic latitudes (Cohen et al. \cite{coh})

The identification was considered as reliable if $LR>2$. This may not be very correct in this case, since we identify radio sources in dense groups of galaxies. For this reason we inspected each identification on the DSS images.

Ninety three of the found 353 radio sources were identified with certain objects in 74 ShCGs. The list of identifications is presented in Table 1. The Shakhbazian designation of the group and the number of the galaxy (as labelled in the original ShCG finding charts), considered to be the radio emitter, is given in the first row of column 1. The RA and Dec (J2000.0) of the galaxy are given in columns 2 and 3 of the first row. In the second row the designation of the identified radio source (column 1), its coordinates (column 2 and 3), and the flux density (column 4) are given.

\placetable{tab1}

In five cases (ShCG 054, 163, 298, 347, and 352) the NVSS radio source is located between two nearby galaxies, and it was not possible to determine which of them is the radio emitter. It could not be excluded, however, that both galaxies are radio emitters. 

Two or more radio emitting galaxies were found in thirteen ShCGs.  

Radio sources found within boundaries of nine ShCGs: 051, 057, 120, 203, 248, 250, 330, 346, and 352, coincide appreciably well with the positions of relatively weak objects, not considered as member galaxies of corresponding groups in the original ShCG lists. These identifications should be considered as tentative. Though the mentioned objects could, however, be members of corresponding groups. Future spectral observations of these weak objects may clarify whether they are really members of the corresponding groups. 

Two hundred sixty radio sources detected within the boundaries of 99 ShCGs were not identified with any visible object. They may be just background radio sources. 

\subsection{Remarks on identifications}

{\it ShCG 001.01}. The central brightest galaxy of S0 type. There is an emission line at 3727 \AA\ in the spectra of this galaxy (Robinson, \& Wampler \cite{row}).

{\it ShCG 003.01}. The central brightest galaxy. The group is known also as VV 153.

{\it ShCG 007.01}. The central brightest galaxy.

{\it ShCG 009.07}. The galaxy is located at the edge of the group.

{\it ShCG 010.01}. The central brightest spiral galaxy. 

{\it ShCG 011.04}. One of the brightest galaxies.

{\it ShCG 016.01}. The brightest galaxy, spiral. As a typical radio galaxy it has two radio emitting lobes on two sides of the galaxy, and a weaker component coinciding with the galaxy itself (Fig. 1a). Radio lobes are at a projected distance of about 10 kpc from the galaxy. The redshift of the group, known also as I Zw 167 and Arp 330, is 0.02913 (Amirkhanian \cite{amir89}). 

The FIR source in TMTST was tentatively identified with galaxy No. 3, which according to Stoll et al. (\cite{st96}) is an HII region. The consideration of Fig. 2 in TMTST shows that both galaxies, No. 3 and also No. 1 with radio emission, may be FIR emitters. FIR observations with better angular resolution may clarify the situation.
 
{\it ShCG 018.02}. One of the bright galaxies.

{\it ShCG 021.01}. The dominant galaxy of the group with $16\fmag76$. Two other radio sources, FIRST J234646.6-014417 and FIRST J234647.6-014414 seem to be ejected from the brightest galaxy (Fig. 1b).

{\it ShCG 023.02}. One of the bright galaxies.

{\it ShCG 024.02}. The brightest galaxy, $19\fmag36$. 

{\it ShCG 029.03}. The brightest galaxy.

{\it ShCG 033.03}. The 2-d by brightness ($18\fmag35$) galaxy in the group. It is a spiral.

{\it ShCG 040.01}. The dominant cD galaxy (Struble \& Rood \cite{stro87}). The group is known also Abell cluster A0193, z=0.0498 (Struble \& Rood \cite{stro91}). 

{\it ShCG 041.01}. The brightest galaxy of the group.

{\it ShCG 042.11}. A weak galaxy, membership to the group should be proved spectroscopically. The brightest galaxy No. 2, of possibly S0 type, is located nearby. 

{\it ShCG 051.01}. The brightest galaxy, located at the periphery of the group.

{\it ShCG 051.Anon}. A weak object that could possibly be a member of the group. It is not excluded, however, that the radio source may be just a projected one.

{\it ShCG 053.01}. The brightest galaxy of the group, that is known also as Abell cluster A1050.

{\it ShCG 053.04}. One of the bright galaxies.

{\it ShCG 054.06}. One of galaxies of the group. Its membership to the group should be proved by spectral observations. The group is known also as Abell cluster A1067.

{\it ShCG 054.09/10}. A pair weak of galaxies.

{\it ShCG 057.01}. The brightest galaxy.

{\it ShCG 057.02}. Probable identification.

{\it ShCG 062.01}. The brightest galaxy, located at the periphery of the group. More stronger radio source FIRST J112552.1+382153 is located very close to the first one. It may, probably be ejection from the galaxy ShCG 062.01, or a projected source. 

{\it ShCG 065.07}. A galaxy located at the periphery of the group, known also as Abell 1284. The FIR source found here is the galaxy No. 27 in the southern part of the group (TMTST). 

{\it ShCG 074.08}. Relatively weak galaxy, that is very close to the brighter galaxy No. 3.

The FIR source was identified with galaxy No. 9 (TMTST). The center of the FIR source is located almost between galaxies No. 8 and 9. Hence both galaxies may be FIR emitters. 

{\it ShCG 083.01}. The brightest galaxy, $18\fmag02$. 

{\it ShCG 104.03}.  Galaxy is located at about $10\arcsec$ to the south from the brightest galaxy No. 2. The latter was identified as the FIR source (TMTST). It is not excluded, however, that just galaxy No. 3 may really be a FIR emitter.

{\it ShCG 120.Anon}. The central brightest radio source FIRST J110431.2+355157 coincides with a weak object between galaxies No. 4 and 6, located at the periphery of the group. Two weaker radio sources FIRST J110432.6+355212 and FIRST J110433.1+355222 seem to be radio lobes ejected from the central object (Fig. 1c). The group is known also as a cluster A1151. 

The belonging of the object to the group should be checked by spectral observations. The FIR emitter, located just here, was considered as an uncertain identification in MTST. Apparently the same object may really be a FIR emitter.

{\it ShCG 141.01}. The brightest galaxy, $18\fmag93$. 

{\it ShCG 149.01}. One of the brightest galaxies in the central region of the group.

{\it ShCG 163.01/03}. Brightest galaxies (probably interacting) in the center of the group. 

{\it ShCG 166.02}. One of the bright galaxies of the group, known also as a cluster A2247.

{\it ShCG 168.06}. One of the bright galaxies of the group. The FIR source was identified with the brightest galaxy No. 1 (TMTST). 

{\it ShCG 177.01}. The brightest galaxy. 

{\it ShCG 181.04}. The galaxy No. 4 is located nearby to the brightest galaxy No. 1.

{\it ShCG 182.01}. The brightest galaxy, located at the end of the elongated group.

{\it ShCG 186.01}. One of three, probably interacting brightest galaxies. 

{\it ShCG 194.01}. The dominant galaxy. 

{\it ShCG 199.01}. The dominant galaxy. 

{\it ShCG 199.03}. The second by brightness. 

{\it ShCG 202.03}. The brightest galaxy. 

{\it ShCG 203.Anon}. A pair of relatively strong radio sources is identified with very weak object in the north part of the group (Fig. 1d). The pair was not considered as a member of the group in the original ShCG lists. Spectral observations are needed to clarify the case.

{\it ShCG 205.03}. One of the brightest galaxies.

{\it ShCG 209.01}. The brightest galaxy located in the center of the group. 

{\it ShCG 219.01}. Radio source FIRST J145233.5+275751 is identified with the brightest galaxy. It seem to be in interaction with galaxy No. 2. Galaxy No. 1 has halo or spiral arms. The group is known also as a cluster A1984.

Two other radio sources, FIRST J145231.6+275807 and FIRST J145234.2+275751, seem to be ejected from galaxy No. 1 (Fig. 1e). This galaxy and the galaxy No. 1 are  in interaction.

{\it ShCG 234.04}. Relatively weak galaxy, located in the central part of the group. Its membership to the group should be proved by spectral observations.

{\it ShCG 248.04}. The the brightest galaxy of the group. In TMTST the FIR source was identified with the same galaxy.

{\it ShCG 248.Anon}. Relatively bright galaxy located at about $1\arcmin$ to the west from the group. It may be a member of the same group. The consideration of the isophotes of the FIR source (TMTST) shows that both this galaxy and the galaxy No. 4 of the group may be FIR emitters.

{\it ShCG 250.Anon}. Weak object that could possibly be a member of the group. It is not excluded, however, that the radio source may be just a projected one to this group.

{\it ShCG 273.06}. Relatively bright galaxy, what, probably, is in interaction with galaxy No. 3. The FIR source was identified with galaxy No. 3 (TMTST). The finding of the radio emission from this galaxy suggests that just it may be the FIR source. The errors of the IRAS positional measurements in this case are $18\arcsec\times6\arcsec$.

{\it ShCG 279.04}. Relatively weak galaxy, $17\fmag75$. Probable identification.

{\it ShCG 282.04}. Galaxy of S0 type, that is apparently interacting with the brightest galaxy No. 1 of Sbc type.

{\it ShCG 289.03}. E type galaxy, that apparently interacts with galaxies No. 1 (of E/S0 type) and 4. Galaxy No. 1 is the brightest galaxy, $17\fmag55$, and galaxy No. 3 is the third, being a little bit weaker, $17\fmag97$.

{\it ShCG 298.02/06}. A pair of apparently interacting  bright galaxies. The stellar magnitudes are equal to $18\fmag71$ and $19\fmag55$ respectively.

{\it ShCG 303.03}. Bright galaxy in the center of the group, apparently is interacting with another bright galaxy No. 4. 

{\it ShCG 309.07}. The brightest galaxy, $17\fmag49$. It seems to be interacting with galaxy No. 8.

{\it ShCG 312.10}. The dominant galaxy. It seems to be in interaction with galaxy No. 9, and has a couple of dwarf satellites.

{\it ShCG 317.01}. The FIRST Survey shows very complex structure (Fig. 1f) of the radio source identified with the brightest galaxy of the group ($15\fmag72$). It has several components connected with bridges. The overall size of the radio complex is about 100 kpc. The redshift of the group is 0.0434 (Tiersch et al., in preparation). 

{\it ShCG 320.09}. One the bright galaxies, $17\fmag88$.

{\it ShCG 329.03}. The brightest galaxy of the group, $17.\fmag96$.
 
{\it ShCG 330.Anon}. A weak object that could possibly be a member of the group. It is not excluded, however, that the radio source may be just a projected one.

{\it ShCG 331.07}. The brightest galaxy, $16\fmag06$. The FIR source was identified with the same galaxy (TMTST).  

{\it ShCG 335.02}. The 2-d by brightness galaxy, $18\fmag27$.

{\it ShCG 339.01}. One of the three brightest galaxies. 

{\it ShCG 340b.05}. One of the brightest galaxies.

{\it ShCG 344.07}. Galaxy of Sab type. Seems to be in interaction with another, relatively weak galaxy.

The FIR source was identified in TMTST with the brightest galaxy No. 1 of S0 type. The reconsideration of the FIR isophotes here with taking into account the errors of the IRAS positional measurements in this case ($40\arcsec\times23\arcsec$) allows to suggest that galaxy No. 7 may also be a FIR emitter.  

{\it ShCG 346.Anon}. A weak object, that could possibly be a member of the group. It is not excluded, however, that the radio source may be just a projected one.

{\it ShCG 347.01}. The brightest galaxy, located at the end of the elongated group.

{\it ShCG 347.03/04}. The radio source is located between galaxies No. 3 and 4, and could be identified with one of them.

{\it ShCG 348.02}. One of two brightest galaxies in the group. 

{\it ShCG 351.06}. The dominant spiral galaxy (UGC 06212). It is located at the periphery of the group. In TMTST the FIR source was identified with the same galaxy.

{\it ShCG 352.Anon}. The position of the radio source coincides well with the weak object at about $17\arcsec$ to the S-W from the brightest galaxy No. 1.

{\it ShCG 359.01}. The brightest galaxy.

{\it ShCG 360.01}. The brightest galaxy of S0 type. This very compact group, is known also as the cluster A2113.

{\it ShCG 362.01/04}. The radio source is located at about $12\arcsec$ south of galaxy No.4, and could be identified with it, or with the brightest galaxy No. 1, located nearby. These two galaxies seems to be interacting. The group is known also as III Zw 108.

{\it ShCG 371.02}. The group is very dense. The radio source is identified with one of the three central bright galaxies of the group, Nos. 2, 3 or 4. The latter is the brightest in the group. In TMTST the FIR source was tentatively identified with galaxy No. 2. However, the optical spectrum of these galaxy is a normal one with absorption lines, while the spectrum of galaxy No. 4 has emission lines. (The results of the spectral observations of this group will be published elsewhere). On the basis of spectral data we assume that galaxy No. 4 may also be a FIR emitter. 

{\it ShCG 372.03}. One of the bright galaxies.

{\it ShCG 376.04}. The dominant galaxy. The FIR source was identified with the same galaxy (TMTST). Spectral observations of this group (the results of which will be published elsewhere) show that this galaxy, and also some others in the group, have emission lines.

\subsection{The optical rank of the radio emitting galaxy}

A radio emission, as we already mentioned, is most often observed from the optically first-ranked galaxies or other brightest members of groups of galaxies (Tovmassian \& Shakhbazian \cite{tovsh81}, Menon \cite{men92}). We identified most of radio sources detected in ShCGs (64) with the brightest or one of the bright galaxies in corresponding groups. Only in nineteen cases the identified objects are not the brightest members. If the found regularity (Tovmassian \& Shakhbazian \cite{tovsh81}, Menon \cite{men92}) holds in the case of ShCGs as well, then some of the latter nineteen identifications with weaker members of groups may not be correct. The membership of these objects to the corresponding groups should be checked by spectral observations.

\subsection{The structure of radio sources}

Due to large distances of ShCGs from us the angular sizes of member galaxies are generally small enough. For this reason the angular resolution of the FIRST ($5\arcsec$) and, especially of the NVSS ($45\arcsec$) does not allow in most cases to distinguish whether the observed radio radiation is emitted from the nuclear region of the galaxy, or from its disk. Hence we assumed that the measured flux refers to the disc.

However, in some cases the high angular resolution of the FIRST Survey allowed, to resolve detected radio sources. Radio sources, identified with ShCG 016.01, 021.01, 219.02, and 317.01 have composite structure (Fig. 1). Some of them consist of two lobes located diametrically in two sides of the parent galaxy, that is characteristic to classical radio galaxies. In the case of ShCG 016.01 and 219.01 the radio sources are of FR II type. Radio sources identified with ShCG 120.Anon and 203.Anon also seem to consist of two components. In ShCG 016.01, 120.Anon, and 317.01 the radio emission of the central galaxy itself is also observed. The radio source in ShCG 317 is a very complex one (Fig. 1).

\placefigure{fig1}

\subsection{Radio spectra}

Radio sources detected within ShCG areas were cross-identified with sources of the CATS (astrophysical CATalogs Support system) database (Verkhodanov et al. \cite{verch}), that unifies 200 radio astronomical catalogs, including the Texas catalog at 365 MHz (Douglas et al. \cite{doug}), 6C at 151 MHz (Baldwin et el. \cite{bald}), high sensitivity WENS at 327 MHz (Rengelink et al., \cite{reng}), and others. We used the task {\it match} in the identification circle of a $90\arcsec$ radius. We found that twenty two of the detected radio sources have been observed at least at two frequencies. The spectra of these radio sources are presented in Fig. 2. The spectral indices of these sources were determined. For determination of spectral indices we used the least square methods to fit the obtained data sample for construction of the spectra.

\placefigure{fig2}

The spectral indices are presented in Tab. 2. Most of the spectra have normal slopes. The radio source, identified with ShCG 248.04 have very unusual, too steep spectrum. It is possible that the flux density of this source at 365  MHz is overestimated. The spectra of three sources, ShCG 041.01,  051.04 and 163.01/03 are inverted. 

\placetable{table2}

\subsection{Radio luminosities} 

Redshifts of only a handful ShCGs have been measured until recently (Robinson \& Wampler \cite{row}; Arp et al. \cite{arp73}; Mirzoyan et al. \cite{mirz}; Amirkhanian \& Egikian \cite{amir87}; Amirkhanian \cite{amir89}; Kodaira et al. \cite{kod88}, 1990; Kodaira \& Sekiguchi \cite{kods91}; Lynds et al. \cite{lynd}; del Olmo \& Moles \cite{delolm}). The redshifts of only 37 ShCGs with detected radio sources are known. Most of these redshifts are yet unpublished (Tiersch et al., \cite{ti99}). We derived the radio luminosities of these sources at 1.4 GHz by assuming $H=50$ $km~s^{-1} Mpc^{-1}$ and $q_{0}=0.5$. The mean redshift of the group members, if available, was used in calculations. The derived radio luminosities are presented in Table 3. 

\placetable{tab3}

We compared radio luminosities of galaxies in ShCGs with that of in HCGs (Fig. 3). For drawing Fig. 3 we used total fluxes of 56 HCG spiral galaxies (Menon \cite{men95}) at 1415 MHz, and also 34 more HCG galaxies identified with the NVSS and FIRST radio sources by us. The list of the latter galaxies is presented in Table 4. 

\placetable{tab4}

\placefigure{fig3}

The consideration of Fig. 3 shows that radio sources in ShCGs are more powerful.  Indeed, the radio luminosities of more than half of HCGs located at redshifts $<0.05$ are less than 22.0 $WHz^{-1}$, while only one out of eleven ShCGs at the same distances has such low radio luminosity. It is seen also that most of powerful ShCG radio sources are located at larger distances, where no HCGs were found.

\subsection{The comparison of the radio and FIR emitting properties of ShCGs and HCGs}

Since redshifts are known for only a limited sample of ShCGs, it is not yet possible to study the radio luminosity function of ShCG galaxies. Such study may be done after compilition of the program initieted by Tiersch et al. (\cite{ti99}) on the spectral study of ShCGs. The available data allow, anyhow, to compare the radio and FIR emission of HCGs and ShCGs. 

For comparison of the radio and FIR emission abilities of ShCGs and HCGs we draw the graph log $F_{60}$-log $F_{1.4}$ (Fig. 4). The 60~$\mu$m IRAS band was by far the most sensitive for the detection of extragalactic objects (see, for example, \cite{hick89}, \cite{surmaz}).

For star forming galaxies a very close connection between two apparently unrelated physical mechanisms, the thermal emission from a dust and the synchrotron radio emission from relitivistic electrons, was found (Dickey \& Salpeter \cite{dick}, Helou, Sofier \& Rowan-Robinson \cite{hel85}, Hickson et al. \cite{hick89}, Helou \& Bicay (\cite{heb93}). It was shown that the ratio of the FIR and radio fluxes of starburst galaxies is almost constant. 

If the ratio of the FIR and radio fluxes of galaxies in the considered CGs is also constant, as it is in the star forming galaxies, then due to different distances from us, and also due to differences in the emitted fluxes, the CG galaxies should be distributed on Fig. 4 along a diagonal line.

For drawing Fig. 4 we used:

- thirty four spiral HCG galaxies from Menon's list (\cite{men95}), and also twelve HCG galaxies from Tab. 4 (this paper), the FIR emission of which at 60 $\mu$m was mesured by Allam et al. (\cite{al}). Nine galaxies from Table 4 also are spirals. 

- eleven ShCG galaxies with detected radio and FIR emission (TMTST). These are galaxies in groups ShCG 016, 074, 104, 120, 168, 248, 273, 331, 344, 371, and 376.

- Markarian-starburst (SB) and Markarian-Seyfert (Sy) galaxies. The FIR (at 60 $\mu$m) and radio fluxes of Markarian galaxies are taken from Bicay et al. (\cite{bick95}).

\placefigure{fig4}

The high accuracies of radio positional measurements allowed to identify with high confidence the detected radio source with a certain galaxy in the corresponding CG. The situation is not the same in the case of the IRAS FIR observations. Absolute positions provided by the IRAS are accurate up to  $6\arcsec$ (within $\pm3\sigma$) in the in-scan and $\sim25\arcsec$ in the cross-scan directions respectively. In the case of HCGs, that are nearer to us and have relatively larger angular dimensions in comparison with ShCGs, in most cases (37 out of 47) it was possible to identify the detected FIR source with a certain galaxy in the corresponding group (Allam et al. \cite{al}). The FIR sources detected in ShCGs (TMTST) are in general weaker than those in HCGs. As a result their positional measurement accuracies reach usually values of about $10\arcsec$ to $30\arcsec$ in the in-scan and the cross-scan directions respectively. For this reason, and also for the relatively smaller angular sizes of ShCGs it was not possible to determine with certainty which galaxy in a dense group is the FIR emitter (TMTST). As a probable one the brightest member of the group was usually mentioned. For construction of Fig. 4 we attributed the measured FIR flux eather to a single galaxy or to two nearby galaxies (see Subsection 3.2). This uncertainty does not however influence the made conclusions.

The consideration of Fig. 4 shows that most of HCG galaxies (30 out 47, i.e. $\sim64\%$) are located along and somewhat lower of the arbitrarily drawn diagonal dashed line. These galaxies thus obey the correlation between the thermal emission from a dust and the synchrotron radio emission from relitivistic electrons found for star forming galaxies, and hence they also are SB galaxies. Galaxies that are lower of the dashed line apparently have relatively stronger FIR emission. Along the same dashed line are distributed, as it was expected, also most of Markarian SB galaxies. 

Seventeen HCG galaxies ($\sim36\%$) are higher of the dashed line. It means that they have stronger radio emission than SB galaxies. It is remarkable that two of these galaxies, HCG 92c and HCG 96a, are Seyferts, and three others, HCG 56b, HCG 68a and HCG 68b, are E and S0 type galaxies. 

The situation is different for ShCG galaxies. Nine of them (82\%) are located above the arbitrarily drawn diagonal line. Above this line are located also twelve out of thirteen ($\sim92\%$) Markarian-Seyfert galaxies. Hence most of ShCG galaxies are not SB galaxies. If the identification of the FIR source with a radio emitting galaxy is not correct then the galaxy with detected radio emission and thus with smaller FIR emission would move on Fig. 4 to the left and hence would be located even higher of the dashed line. The blending of a few probable FIR sources in dense groups would have the same effect.

It is worth to note that those HCG and ShCG galaxies for which either FIR or radio emission was detected would also have different locations on the log$F_{60}$-log$F_{1.4}$ graph. 

There are 73 ShCG galaxies, fluxes of which at 1.4 GHz exceed 1 Jy, but only upper limits of fluxes at 60 $\mu$m were determined (TMTST). These galaxies would apparently be located on Fig. 4 higher of the dashed line. At the same time there are only fifteen ShCG galaxies with determined FIR fluxes (TMTST), and radio fluxes lower than 1 Jy detection limit (this paper). The latter would be located lower of the dashed line. In the case of HCG galaxies the situation is vice versa. There are 34 HCG galaxies with 1.4 GHz fluxes exceeding the 1 mJy limit (Menon \cite{men95}, and present paper), and the FIR fluxes lower than the limiting value (Allam et al. \cite{al}). These galaxies would be located on Fig. 4 upper of the dashed line. Much more HCG galaxies, 65, with measured FIR fluxes and upper limits of radio fluxes, would be located below the dashed line, in the lower part of Fig. 4. Thus in the case of galaxies with eather one of the fluxes (eather 60 $\mu$m or 1.4 GHz) measured, we see the same trend: most ShCG galaxies would be located upper of the dashed line on Fig. 4. Meanwhile HCG galaxies would be located mainly below this line.

Sulentic \& De Mello Rabaca (\cite{suldmr}) claimed that the FIR sources in HCGs are likely the combined contribution of two or more members. If to assume that in all Hickson's groups with detected FIR emission, the FIR emitters are in reality two galaxies with equal fluxes, then the corresponding points on Fig. 4 should move to the left by 0.3, and still they would be located lower than ShCG galaxies. Moreover, if the same is valid also for the more dense ShCGs, then the corresponding positions of ShCG galaxies on Fig. 4 should also move to the left, and the found trend would certainly not be altered. 

Hence, one may conclude that galaxies in ShCGs are relatively stronger radio emitters, while HCG galaxies are stronger FIR emitters. It means that physical conditions in ShCGs are somehow favorable for triggering the AGNs with relatively strong synchrotron emission of relativistic electrons, while in HCGs the conditions are favorable for formation of SB galaxies with relatively strong thermal dust emission. What could be the reason for such difference?  

CGs are generally very dense formations. According to N-body simulations (Carnevali et al. \cite{carn}, Barnes \cite{barn85}, \cite{barn89}, Mamon \cite{mam90}, Zheng et al. \cite{zheng93}) the member galaxies at high density environments of CGs should interact and merge in one large galaxy in about $10^{8}$ years. For this reason the very existence of CGs has been questioned by some authors (Walke \& Mamon \cite{wam}, Mamon \cite{mam86}, \cite{mam95}, Hernquist et al. \cite{hern}). Meanwhile Hickson \& Rood (\cite{hickr}), Mendes de Oliveira \& Giraud (\cite{meg}), Mendes de Oliveira (\cite{me}), Oleak et al. (\cite{oleak95}) and recently Tovmassian et al. (\cite{tov98c}) presented firm evidences on the reality of CGs. For explaining the existence of CGs Governato et al. (\cite{gov}) proposed a second generation merger scenario, according to which CGs permanently aggregate new members from their surrounding areas. Such scenario could be valid since, according to Rood \& Williams (\cite{roodw}), Vennik et al. (\cite{ven}) and Ramella et al. (\cite{ram}), most of HCGs are associated with loose groups of galaxies.

ShCGs and HCGs differ from each other by the number of members and by morphological content. If HCGs contain generally four-five members, ShCGs have up to fifteen members. The relative number of E or S0 type galaxies in HCGs compose only $\sim~50\%$ of all members (Hickson et al. \cite{hick89}), while ShGGs are more rich in early type galaxies. About 77\% member galaxies in the latter are of E and S0 type (Tiersch et al. \cite{ti96a}). For this reason ShCGs are considered as more evolved systems in comparison with HCGs. 

Since almost half of HCG galaxies are spirals with sufficient amount of gas and dust, the gravitational interaction between them and with a suggested newcomer galaxy may trigger starburst processes in interacting galaxies. The FIR emission is characteristic to such events. Meanwhile in the more evolved ShCGs most galaxies are of E and S0 type, and the number of spirals is relatively small. The gas was pushed out from them during previous interaction processes between galaxies, and filled the intergalactic space. The spirals here should have shed also their dust content. The presence of intergalactic gas in CGs was proved by the detection of diffuse X-ray emission (Ebeling et al. \cite{eb}, Pildis et al. \cite{pild}, Saracco \& Ciliegi \cite{sarc}, Tiersch et al. \cite{ti96b}) from some of the groups. Due to the interaction with a newcomer galaxy to the group this intergalactic gas may be falling in the form of cooling flows directly into the center of the preferentially dominant early type galaxy in the group. In the result of this an active nucleus of the galaxy with sufficiently strong radio emission may be formed. Due to small amount or even absence of dust in such galaxies their FIR emission would be very weak or even completely absent.

\section{CONCLUSIONS}

Three hundred fifty three NVSS (Condon et al. \cite{con98}) radio sources are found within boundaries of 179 ShCGs. Sixty sources were registered also in the FIRST (White et al. \cite{white}) survey, of which seven sources - only in the FIRST.

Ninety three of the found radio sources are identified with corresponding galaxies in 74 out of 366 ShCGs. 

The complex structure of radio sources in ShCG 016, 021, 120, 203, 219, and 317 is found.

Radio spectra of 22 radio sources are constructed.

It is shown that ShCG galaxies have in general higher radio luminosities than galaxies in HCGs.

The comparison of the radio and FIR fluxes of galaxies in HCGs and ShCGs showed that the latter are more stronger radio emitters, while HCG galaxies are generally more stronger FIR emitters. It is suggested that the reason for this may be that ShCGs are more evolved in comparison with HCGs, and galaxies in them do not have enough dust for attributing the FIR emission. On the other hand the conditions in ShCGs are more favorable for formation of AGN with relatively stronger radio emission.

\acknowledgements

HMT acknowledge the support of the Deutsche Forschungsgemeinschaft DFG project No. 444Mex 112/2/98. VHCh was partially supported by CONACYT research grant 28499-E. OVV is grateful to the Russian Foundation for Basic Researches for the CATS database support (grant No 96-07-89075). HT is grateful to the Government of the Land Brandenburg and the Deutsche Forschungsgemeinschaft DFG project No. TI 215/6-3 for the support of this work.

The authors are indebted to the anonymous referee for critical comments and useful suggestions.

The Digitized Sky Surveys were produced at the Space Telescope Science Institute under U.S. Government grant NAG W-2166. The images of these surveys are based on photographic data obtained using the Oschin Schmidt Telescope on Palomar Mountain and the UK Schmidt Telescope.





\clearpage

{\scriptsize
\begin{planotable}{lllr}
\tablenum{1}
\tablewidth{30pc}
\tablecaption{The list of radio identifications}
\tablehead{
\colhead{{\it ShCG.G-xy}}  & \colhead{R.A.} & \colhead{Dec.} & \colhead{Flux}
\\[1ex]
\colhead{Radio source}   & \colhead{J2000.0} & \colhead{J2000.0} & \colhead{mJy}   }

\startdata
{\it 001.01}           & 10 55 05.7  & +40 27 30     \nl     
NVSS J105506+402726    & 10 55 06.18 & +40 27 26.7   & 3.9  \nl
{\it 003.01}           & 11 15 23.4   & +53 41 23     \nl
FIRST J1115  +534122   & 11 15 23.504 & +53 41 22.28 & 1.32 \nl
{\it 007.01}           & 11 05 53.8   & +39 46 58    \nl
FIRST J110553.7+394654 & 11 05 53.766 & +39 46 54.81 & 39.93   \nl      
{\it 009.07}           & 13 24 01.5 & +19 03 21       \nl
NVSS J132401+190320    & 13 24 01.74 & +19 03 20.2   & 4.2     \nl                  			{\it 010.01}           & 14 10 48.1   & +46 15 58  \nl
FIRST J141448.1+461557 & 14 10 48.176 & +46 15 57.45 & 11.78   \nl              			{\it011.04}           & 14 11 01.7   & +44 42 15         \nl
FIRST J111411+444214   & 14 11 01.826 & +44 42 14.34 & 1.72  \nl
{\it 016.01}           & 16 49 11.3   & +53 25 12        \nl
FIRST J164911.4+532510 & 16 49 11.496 & +53 25 10.91 & 23.43   \nl
FIRST J164910.5+532507 & 16 49 10.398 & +53 25 07.31 & 62.14   \nl
FIRST J164912.5+532514 & 16 49 12.582 & +53 25 14.56 & 53.58    \nl
{\it 018.02}           & 08 53 37.2  & +79 09 17            \nl
NVSS J085339+790915    & 08 53 39.73 & +79 09 15.4 & 3.8     \nl
{\it 021.01}           & 23 46 48.6   & -01 44 16  \nl 
FIRST J234648.6-014416 & 23 46 48.633 & -01 44 16.75 & 64.52   \nl
FIRST J234646.6-014417 & 23 46 46.680 & -01 44 17.41 & 13.70  \nl
FIRST J234647.6-014414 & 23 46 47.641 & -01 44 14.56 &  6.72   \nl
{\it 023.02}           & 16 10 03.2   & +52 14 52          \nl
FIRST J161003+521450   & 16 10 03.560 & +52 14 50.33 & 1.15 \nl
{\it 024.01}           & 23 46 56.2   & -00 52 27        \nl
FIRST J234656.3-005227 & 23 46 56.352 & -00 52 27.21 & 2.19   \nl         				{\it 029.03}           & 16 08 45.1   & +52 26 17 \nl
FIRST J160845.1+522616 & 16 08 45.191 & +52 26 13.1 & 2.48  \nl         			{\it 033.03}           & 01 03 42.7   & -01 08 13   \nl	
FIRST J010342.5-010813 & 01 03 42.521 & -01 08 13.10 & 3.80  \nl                  			{\it 040.01}           & 01 25 07.6  & +08 41 59    \nl				 NVSS J012510+084224    & 01 25 07.87 & +08 41 59.2 & 33.4   \nl     
{\it 041.01}           & 01 29 00.6  & +07 40 40  \nl
NVSS J012900+074042    & 01 29 00.61 & +07 40 42.1 & 36.8   \nl    
{\it 042.11}           & 01 30 44.6  & +07 50 09         \nl
NVSS J013044+075004    & 01 30 44.65 & +07 50 04.4 & 3.3 \nl                   			{\it 051.01}           & 10 30 44.6  & +39 12 45      \nl
FIRST J103044.7+391245 & 10 30 44.619 & +39 12 44.28 & 14.2  \nl
{\it 051.Anon}         & 10 30 44.1 & +39 10 29   \\
FIRST J103044.0+391028 & 10 30 44.196 & +39 10 29.73 & 5.64  \\ 
{\it 053.01}           & 10 36 46.7  & +44 49 48  \nl
FIRST J103646.4+444946 & 10 36 46.496 & +44 49 46.39 & 2.74  \nl                      {\it 053.04}           & 10 36 52.8  & +44 48 21  \nl     
FIRST J103653.0+444818 & 10 36 53.027 & +44 48 18.19 & 35.82  \nl     
{\it 053.14}           & 10 36 45.7  & +44 49 54  \nl     
FIRST J103645.9+444955 & 10 36 45.988 & +44 49 55.16 & 3.86  \nl     
{\it 054.06}           & 10 40 37.2  & +40 12 58 \nl
FIRST J104037.3+401257 & 10 40 37.320 & +40 12 57.92 & 1.16  \nl
{\it 054.09}           & 10 40 27.1  & +40 13 41   \nl
{\it 054.10}           & 10 40 27.0  & +40 13 48    \nl
NVSS J104026+401345    & 10 40 26.92 & +40 13 45.9 & 11.2  \nl    
{\it 057.01}           & 10 45 26.7   & +49 31 08  \nl 
FIRST J104527.2+493106 & 10 45 27.266 & +49 31 06.27 & 20.01 \nl
{\it 057.02}           & 10 45 26.1   & +49 31 25  \nl
FIRST J104526.4+493116 & 10 45 26.492 & +49 31 16.98 & 29.92  \nl    
{\it 062.01}           & 11 25 53.2   & +38 22 04 \nl
FIRST J112553.3+382201 & 11 25 53.305 & +38 22 01.38 & 28.95  \nl    
FIRST J112552.1+382153 & 11 25 52.121 & +38 21 53.30 & 30.32 \nl  
{\it 065.07}           & 11 30 50.6   & +35 04 15 \nl
FIRST J113050.5+350415 & 11 30 50.578 & +35 04 15.03 & 4.79   \nl
{\it 074.08}           & 14 20 57.3   & +43 02 54 \nl
FIRST J142057.5+430250 & 14 20 57.562 & +43 02 50.45 & 2.39  \nl
{\it 083.01}           & 23 26 08.8   & -01 43 30                   \nl	
FIRST J232608.8-014329 & 23 26 08.867 & -01 43 29.83 & 5.849  \nl
{\it 104.03}           & 09 27 13.5   & +52 58 33                   \nl
FIRST J092713+525832   & 09 27 13.561 & +52 58 32.58 & 2.08  \nl
{\it 120.Anon}         & 11 04 31.3   & +35 52 08 \nl   
FIRST J110431.2+355157 & 11 04 31.227 & +35 51 57.70 & 3.72  \nl
FIRST J110432.6+355212 & 11 04 32.605 & +35 52 12.54 & 71.46  \nl
FIRST J110433.1+355222 & 11 04 33.137 & +35 52 22.14 & 9.01   \nl
{\it 141.01}           & 01 04 22.2   & -01 33 27                    \nl	
FIRST J010422.2-013326 & 01 04 22.203 & -01 33 26.54 & 12.67  \nl
{\it 163.03}           & 15 21 03.9   & +75 04 20               \nl
{\it 163.01}           & 15 21 04.8   & +75 04 12               \nl    
NVSS J152105+750421    & 15 21 05.13  & +75 04 21.5 & 7.1  \nl      
{\it 168.06}           & 18 28 03.2   & +83 06 07                 \nl
NVSS J182806+830605    & 18 28 06.27  & +83 06 05.8 & 169.5 \nl      
{\it 177.01}           & 01 57 36.3   & +29 38 56              \nl
NVSS J015736+293853    & 01 57 36.50  & +29 38 53.5 & 43.4  \nl    
{\it 181.04}           & 08 28 00.0   & +28 15 37 \nl
NVSS J082800+285134    & 08 28 00.28  & +28 15 34.1 & 10.0 \nl
{\it 182.01}           & 08 38 23     & +29 45 22  \nl
FIRST J083823.2+294521 & 08 38 23.271 & +29 45 21.67 & 3.92  \nl                      {\it 186.01}           & 09 22 51.2   & +28 55 53 \nl
FIRST J092251.1+285552 & 09 22 51.148 & +28 55 52.89 & 5.17  \nl
{\it 194.01}           & 11 03 06.7   & +27 48 25                   \nl
FIRST J110306+274823   & 11 03 06.800 & +27 48 23.19 & 1.13  \nl
{\it 199.01}           & 11 35 22.7   & +30 43 43 \nl
FIRST J113522.8+304343 & 11 35 22.855 & +30 43 43.14 & 1.45  \nl      
{\it 199.03}           & 11 35 21.3   & +30 42 58 \nl
FIRST J113521.4+304257 & 11 35 21.480 & +30 42 57.20 & 0.77  \nl      
{\it 202.03}           & 12 19 51.8   & +28 25 18 \nl
FIRST J121951.6+282521 & 12 19 51.668 & +28 25 21.51 & 8.00  \nl                      {\it 203.Anon}         & 12 29 02.7   & +27 27 11 \nl
FIRST J122902.3+272702 & 12 29 02.387 & +27 27 02.36 & 179.56   \nl                      FIRST J122902.7+272731 & 12 29 02.770 & +27 27 31.58 & 234.00   \nl  
{\it 205.03}           & 12 35 19.0   & +27 34 41  \nl
NVSS J123519+273438    & 12 35 19.05  & +27 34 38.4 & 3.84   \nl
{\it 209.01}           & 13 10 38.9   & +31 44 20 \nl
FIRST J131039.0+314417 & 13 10 39.094 & +31 44 17.01 & 1.88  \nl                       {\it 218.02}           & 14 33 35.4   & +26 41 54                   \nl
FIRST J143335+264153   & 14 33 35.463 & +26 41 53.91 & 1.65  \nl
{\it 219.01}           & 14 52 33.4   & +27 57 51 \nl
FIRST J145233.5+275751 & 14 52 33.543 & +27 57 51.72 & 31.38   \nl     
FIRST J145231.6+275807 & 14 52 31.680 & +27 58 07.84 & 48.27   \nl
FIRST J145234.2+275751 & 14 52 34.297 & +27 57 51.86 & 40.66  \nl
{\it 234.04}           & 10 48 25.0   & +36 15 46 \nl
FIRST J104825.0+361547 & 10 48 25.004 & +36 15 47.71 & 9.09   \nl      
{\it 248.04}           & 13 12 15.6   & +36 11 10  \nl
FIRST J131216.0+361108 & 13 12 16.040 & +36 11 08.36 & 2.47   \nl                     {\it 248.Anon}         & 13 12 10.0   & +36 11 14   \nl
FIRST J131210.1+361112 & 13 12 10.191 & +36 11 12.45 & 0.91  \nl                     {\it 250.Anon}         & 13 34 47.0   & +33 08 57 \nl
FIRST J133447.0+330857 & 13 34 47.004 & +33 08 57.16 & 7.78  \nl       
{\it 254.03}           & 13 56 19.0   & +35 11 20 \nl
FIRST J135619+351119   & 13 56 19.184 & +35 11 19.85 & 1.35  \nl 
{\it 263.07}           & 00 13 54.6   & -08 38 49 \nl
FIRST J001354-083850   & 00 13 54.851 & -08 38 50.83 & 1.08  \nl
{\it 273.06}           & 02 52 35.5  & -13 06 15 \nl
NVSS J025235-130617    & 02 52 35.85 & -13 06 17.6 & 9.7   \nl   
{\it 282.04}           & 10 52 53.1  & -11 00 21 \nl
NVSS J105252-110020    & 10 52 52.89 & -11 00 20.7 & 75.8  \nl
{\it 289.03}           & 13 58 10.6   & -12 53 03 & \nl
NVSS J135810-125306    & 13 58 10.67  & -12 53 07.0 & 5.8  \nl 
{\it 298.02}           & 22 12 44.5  & -13 40 26\nl
{\it 298.06}           & 22 12 44.8  & -13 40 33  \nl
NVSS J221244-134026    & 22 12 44.84 & -13 40 26.9 & 15.4  \nl     
{\it 303.03}           & 23 17 33.1   & -09 05 32 \nl
FIRST J231733.2-090532 & 23 17 33.297 & -09 05 32.74 & 5.04  \nl                        {\it 309.07}           & 00 51 19.5  & -07 24 39  \nl 
NVSS J005119-072440    & 00 51 19.63 & -07 24 40.1 & 6.3  \nl                       {\it 312.10}           & 01 03 30.0  & -03 32 30  \nl
NVSS J010330-033239    & 01 03 30.97 & -03 32 39.3 & 3.7   \nl
{\it 317.01}           & 02 10 53.3  & -06 33 33  \nl
FIRST J021052.5-063343 & 02 10 52.527 & -06 33 43.22 & 97.0  \nl
FIRST J021053.6-063333 & 02 10 53.615 & -06 33 33.59 & 176.52   \nl
FIRST J021054.2-063344 & 02 10 54.215 & -06 33 44.21 & 96.57  \nl
FIRST J021050.1-063336 & 02 10 50.189 & -06 33 36.38 & 132.83  \nl
{\it 320.09}           & 11 14 46.8  & -06 21 36 \nl	
NVSS J111446-062137    & 11 14 46.59 & -06 21 37.3 & 4.1 \nl 
{\it 329.03}           & 14 37 12.0  & -03 45 51\nl 
NVSS J143712-034547    & 14 37 12.02 & -03 45 47.2 & 8.1 \nl 
{\it 330.Anon}               & 15 14 22.9  & -09 36 07 \nl
NVSS J151423-093603    & 15 14 23.36 & -09 36 03.6 & 6.7  \nl              
{\it 331.07}           & 22 25 26.8  & -02 47 02  \nl
NVSS J222526-024702    & 22 25 26.76 & -02 47 02.4 & 42.6 \nl                       {\it 335.02}           & 23 23 37.4  & -07 24 00   \nl	
NVSS J232337-072357    & 23 23 37.04 & -07 23 57.3 & 4.6  \nl                       {\it 340b.05}          & 00 42 24.6  & +20 22 05  	\nl 
NVSS J004224+202204    & 00 42 24.38 & +20 22 04.5 & 3.3 \nl                      {\it 344.07}           & 08 47 35.8  & +03 42 01  \nl	 
NVSS J084736+034159    & 08 47 36.07 & +03 41 59.9 & 4.5   \nl                        {\it 346.Anon}         & 09 15 12.4  & +05 14 23 \nl			
NVSS J091512+051426    & 09 15 12.20 & +05 14 26.4 & 9.3  \nl                     {\it 347.01}           & 09 17 28.6  & +07 42 31\nl
NVSS J091729+074233    & 09 17 29.10 & +07 42 33.2 & 8.0  \nl      
{\it 347.03}               & 09 17 34.0  & +07 41 10  \\
{\it 347.04}               & 09 17 34.3  & +07 41 22  \\		
NVSS J091729+074233  & 09 17 34.15 & +07 41 16.0 & 18.3  \\      
{\it 348.02}           & 09 26 29.3  & +03 26 17  \nl
NVSS J092629+032617    & 09 26 29.41 & +03 26 17.7 & 10.8  \nl                      {\it 351.06}           & 11 10 24.7  & +04 49 47  \nl
NVSS J111024+044945    & 11 10 24.63 & +04 49 45.3 & 11.93   \nl         
{\it 352.Anon}         & 11 21 31.5  & +02 53 02\nl
NVSS J112131+025303    & 11 21 31.61 & +02 53 03.5 & 5.2   \nl                       {\it 359.01}           & 14 29 54.4  & +18 50 07  \nl
NVSS J142954+185008    & 14 29 54.19 & +18 50 08.6 & 2.6  \nl                       {\it 360.01}           & 15 41 26.5  & +04 43 56 \nl
NVSS J154126+044355    & 15 41 26.54 & +04 43 55.8 & 27.2   \nl            
{\it 362.04}               & 23 32 36.4  & +19 22 27 \\
{\it 362.01}               & 23 32 37.1  & +19 22 33 \\		
NVSS J 233236+192215 & 23 32 36.82 & +19 22 15.4 & 3.6  \\           
{\it 370.03}           & 09 50 20.5  & +23 16 54 \nl
NVSS J095029+231655    & 09 50 20.470 & +23 16 55.54 & 1.26  \nl 
{\it 371.02}           & 11 43 33.1  & +21 53 50  \nl	
NVSS J114333+215406    & 11 43 33.28 & +21 54 06.1 & 4.7  \nl                       {\it 372.03}           & 11 46 49.5  & +24 08 22 \nl
FIRST J114649.5+240821 & 11 46 49.539 & +24 08 21.43 & 2.7  \nl                       {\it 376.04}           & 13 56 35.7  & +23 21 37  \nl
FIRST J135635.7+232135 & 13 56 35.734 & +23 21 35.95 & 4.59                          

\enddata
\end{planotable}

\begin{planotable}{lr}
\tablenum{2}
\tablewidth{8pc}
\tablecaption{The radio spectral indices}
\tablehead{
\colhead{ShCG.G-xy}  &  \colhead{$\alpha$} 
}

\startdata

007.01  &   -0.01 \nl
010.01  &   -1.46  \nl
016.01  &   -0.58 \nl
041.01  &    0.27 \nl
051.01  &    0.79 \nl
053.04  &   -0.12 \nl
054.06  &   -0.75 \nl
054.09/10 & -0.52 \nl
057.01  &   -0.77 \nl
062.01  &   -0.45 \nl
065.07  &   -1.00 \nl
120.Anon &  -0.73 \nl
163.01/03 &  0.26 \nl
168.06   &  -0.72 \nl
177.01   &  -0.63 \nl
182.01   &  -1.01 \nl
203.Anon &  -0.71 \nl
219.01   &  -0.42 \nl
234.04   &  -0.48 \nl
248.04   &  -3.44 \nl
250.Anon &  -1.04 \nl
317.01   & -0.67

\enddata
\end{planotable}

\begin{planotable}{lll}
\tablenum{3}
\tablewidth{14pc}
\tablecaption{The radio luminosities}
\tablehead{
\colhead{ShCG.G-xy}  & \colhead{$z$}  & \colhead{$log P_{1.4}$} 
\\[1ex]
\colhead{}    & \colhead{}        & \colhead{$wHz^{-1}$} 
}

\startdata

001.01 & 0.1168 & 23.36 \\
016.01 & 0.0301 & 23.74 \\
021.01 & 0.0773 & 24.34 \\ 
029.03 & 0.0346 & 22.11 \\
033.03 & 0.0337 & 22.14 \\  
040.01 & 0.0486 & 23.53 \\
041.01 & 0.0900 & 24.11 \\ 
083.01 & 0.0970 & 23.38 \\ 
166.02 & 0.0396 & 24.04 \\
168.06 & 0.1262 & 25.07 \\
181.01 & 0.0917 & 23.56 \\  
202.03 & 0.0262 & 22.38 \\
205.03 & 0.0932 & 23.16 \\
218.02 & 0.0947 & 22.81  \\
248.04 & 0.2712 & 23.90 \\
248.Anon & 0.2712 & 23.46 \\
254.03 & 0.0638 & 22.38 \\
282.04 & 0.1428 & 24.83 \\
289.03 & 0.0706 & 23.10 \\   
298.02/06 & 0.1692 & 24.29 \\  
309.07 & 0.0892 & 23.34 \\
312.10 & 0.0733 & 22.94 \\
317.01 & 0.0434 & 24.61 \\
330.Anon & 0.1078 & 23.53 \\
331.07 & 0.0534 & 23.72 \\
335.02 & 0.0875 & 23.18 \\
340b.05& 0.1045 & 23.19 \\
344.07 & 0.0774 & 23.03 \\
346.Anon & 0.1349 & 23.87 \\
348.02 & 0.0894 & 23.57 \\
351.06 & 0.0290 & 22.64 \\
352.Anon & 0.0490 & 22.73 \\
359.01 & 0.0328 & 22.08 \\
360.01 & 0.1082 & 24.14 \\
362.01/04 & 0.02291 & 21.91 \\
371.02 & 0.1301 & 23.55 \\
376.04 & 0.0660 & 22.93 

\enddata
\end{planotable}

\begin{planotable}{lr}
\tablenum{4}
\tablewidth{16pc}
\tablecaption{The list of HCG radio identifications}
\tablehead{
\colhead{{\it HCG.G-xy}}  & \colhead{Morph. type}
\\[1ex]
\colhead{Radio source}   & \colhead{$F_{1.4}$, mJy}
}

\startdata

{\it 2b}  & cI		\nl
NVSS B002843+081155 & 13.4  \nl
{\it 4a}  & Sc                  \nl
NVSS B003143-214248 & 43.5  \nl
{\it 5a} & Sab                   \nl
NVSS B003619+064718 &  5.7 \nl
{\it 8d}  &  S0                   \nl
NVSS B004656+231800 & 4.1   \nl
{\it 15d}  & E2                  \nl
NVSS B020502+015639 & 5.1   \nl
{\it 21a}  & Sc                 \nl
NVSS B024258-175503 & 4.7	\nl
{\it 21b}  & Sab                 \nl
NVSS B024316-175400 & 2.9	\nl
{\it 25c}  &  Sb               \nl
FIRST J032043.2-010008 & 5.41   \nl
{\it 28b}  &  E5               \nl
NVSS B042457-102607 & 4.4   \nl
{\it 37a}  & E7              \nl
FIRST J091339.4+295934 & 25.58   \nl
{\it 46a}  &  E3              \nl
NVSS B101924+180522 & 4.5  \nl
{\it 48b}  & Sc                \nl
NVSS B103527-265140 & 15.5 \nl
{\it 51c}  &  S0             \nl
FIRST J112230.0+241646 & 7.44  \nl
{\it 53a}  &  SBbc              \nl
NVSS B112613+210419  & 21.7   \nl
{\it 56b}  &  SB0                    \nl
FIRST J113240.2+525701 & 25.81     \nl
{\it 56d}  &  S0                 \nl
FIRST J113235.2+525650 & 2.76     \nl
{\it 58c}  &  SB0a                 \nl
NVSS B113918+103451   & 3.3   \nl
{\it 60a}  &  E2                     \nl
FIRST J120307.2+514030  & 50.89   \nl
{\it 61a}  &  S0a                  \nl
FIRST J121218.8+291046 & 1.45      \nl
{\it 61c}  &  Sbc                   \nl
FIRST J121231.0+291006 & 37.01      \nl
{\it 62a}  & E3                        \nl
NVSS B125029-085604 & 5.4        \nl
{\it 64d}  &  S0                      \nl
NVSS B132306-033526  &  2.3      \nl
{\it 65a}  &  E3                  \nl
NVSS B132703-291520  & 3.5   \nl
{\it 65c}  &  E2                \nl
NVSS B132705-291357  & 3.8   \nl
{\it 68a}  &  S0               \nl
FIRST J135326.6+401658 & 38.30  \nl
{\it 68b}  &  E2                \nl
FIRST J135326.7+401809  & 7.99 \nl
{\it 71a}  &  SBc                 \nl
FIRST J141057.2+252949 & 3.21   \nl
{\it 71b}  & Sb                   \nl
FIRST J141102.5+253110 & 8.40  \nl
{\it 74a}  & E1                   \nl
NVSS B151710+210435  & 16.3   \nl
{\it 78a}  & SBb                 \nl
NVSS B154805+682219   & 7.0  \nl
{\it 79a}  & E0                   \nl
NVSS B155659+205344 & 10.2   \nl
{\it 84a}  & E2          \nl
NVSS B164644+775541  & 21.2  \nl
{\it 86a}  & E2               \nl
NVSS B194859-305716  & 19.6  \nl
{\it 86b}  &  E2                \nl
NVSS B194849-305644  & 8.1  \nl
{\it 94a}  & E1                     \nl
NVSS B231444+182606  & 30.1

\enddata
\end{planotable}
}

\clearpage

\begin{figure}[htb]
\setcounter{figure}{0}
\caption{Radio sources with complex structure. Below each radio image from the FIRST the optical image from the DSS of the corresponding field in the same scale is presented. The sizes of images of ShCG 016.01, 021.01, 120.Anon, and 203.Anon are equal to $1\arcmin\times1\arcmin$. The size of the field of ShCG 219.01 is $1\farcmin5\times1\farcmin5$, and that of ShCG 317.01 is $2\farcmin5\times2\farcmin5$.}
\epsfxsize=15cm\epsfbox{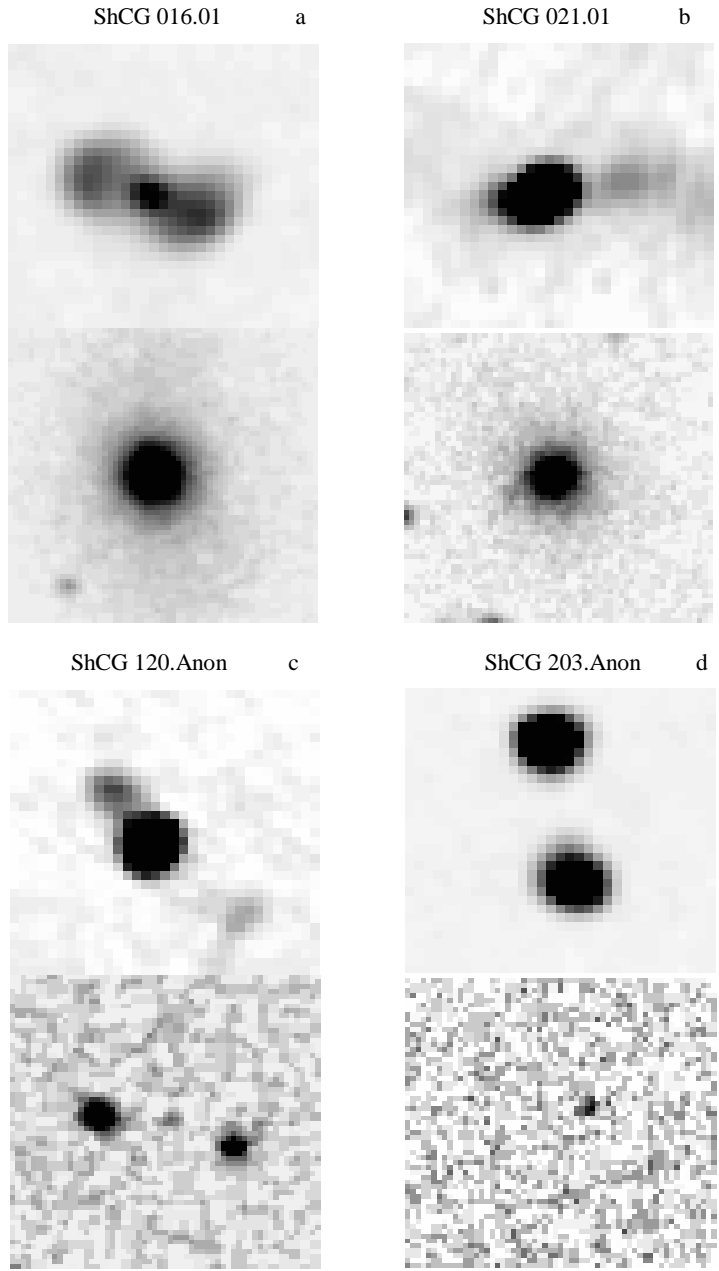}
\end{figure}

\begin{figure}[htb]
\setcounter{figure}{0}
\caption{Continued}
\epsfxsize=15cm\epsfbox{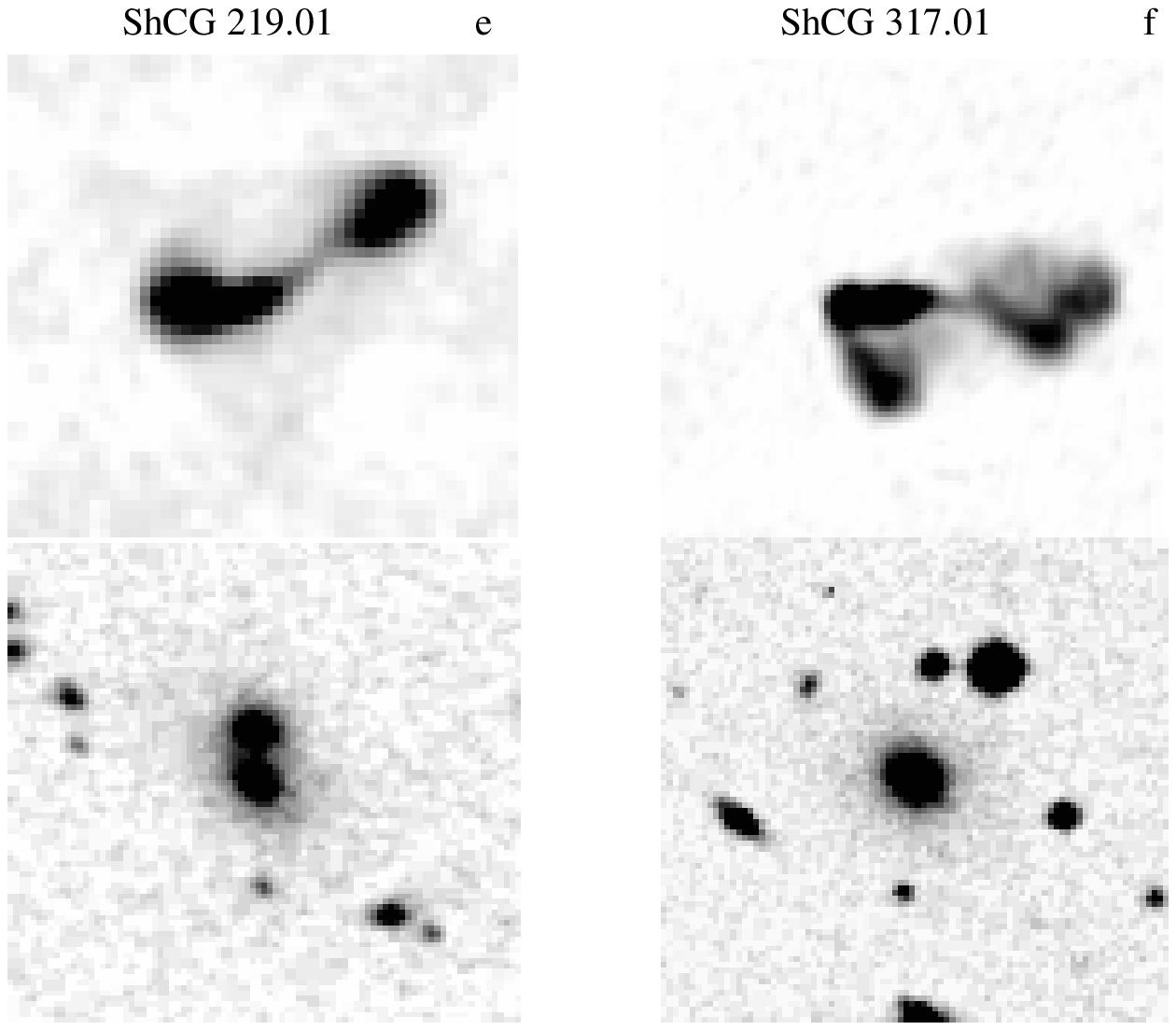}
\end{figure}

\begin{figure}[htb]
\setcounter{figure}{1}
\caption{Radio spectra of some radio sources.}
\epsfxsize=15cm\epsfbox{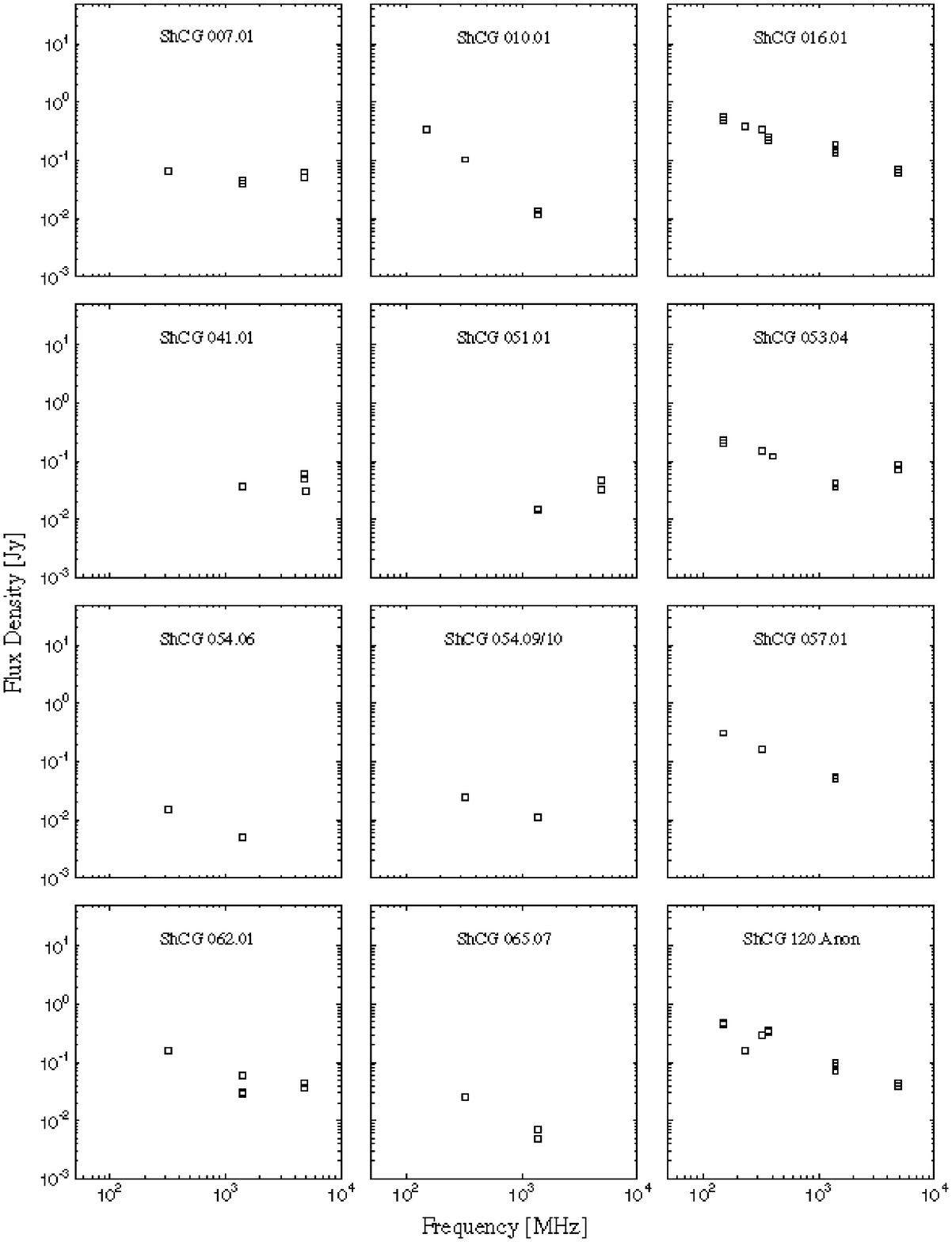}
\end{figure}

\begin{figure}[htb]
\setcounter{figure}{1}
\caption{Continued}
\epsfxsize=15cm\epsfbox{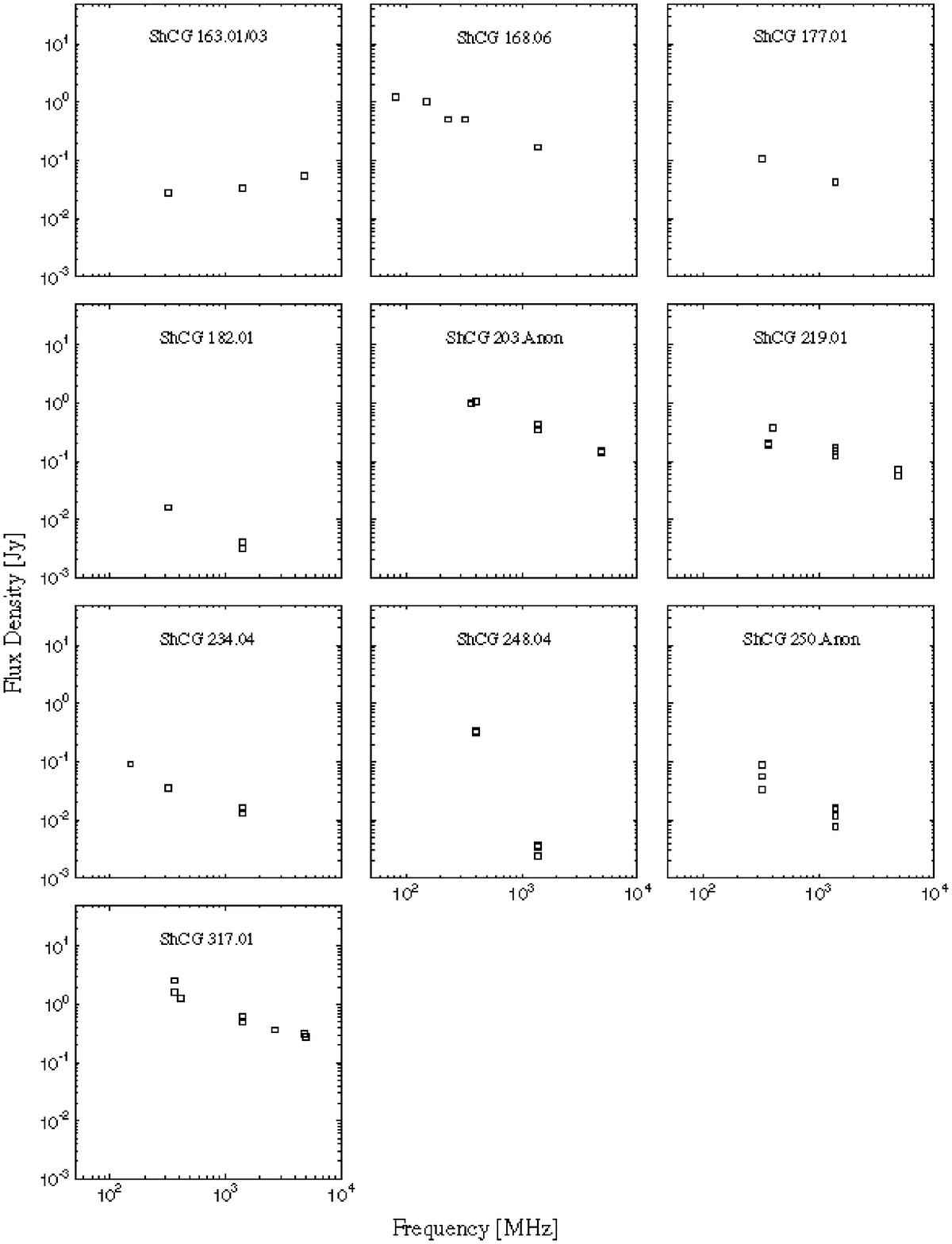}
\end{figure}

\begin{figure}[htb]
\setcounter{figure}{2}
\caption{The dependence of radio luminosity on redshift for HCGs and ShCGs.}
\epsfxsize=15cm\epsfbox{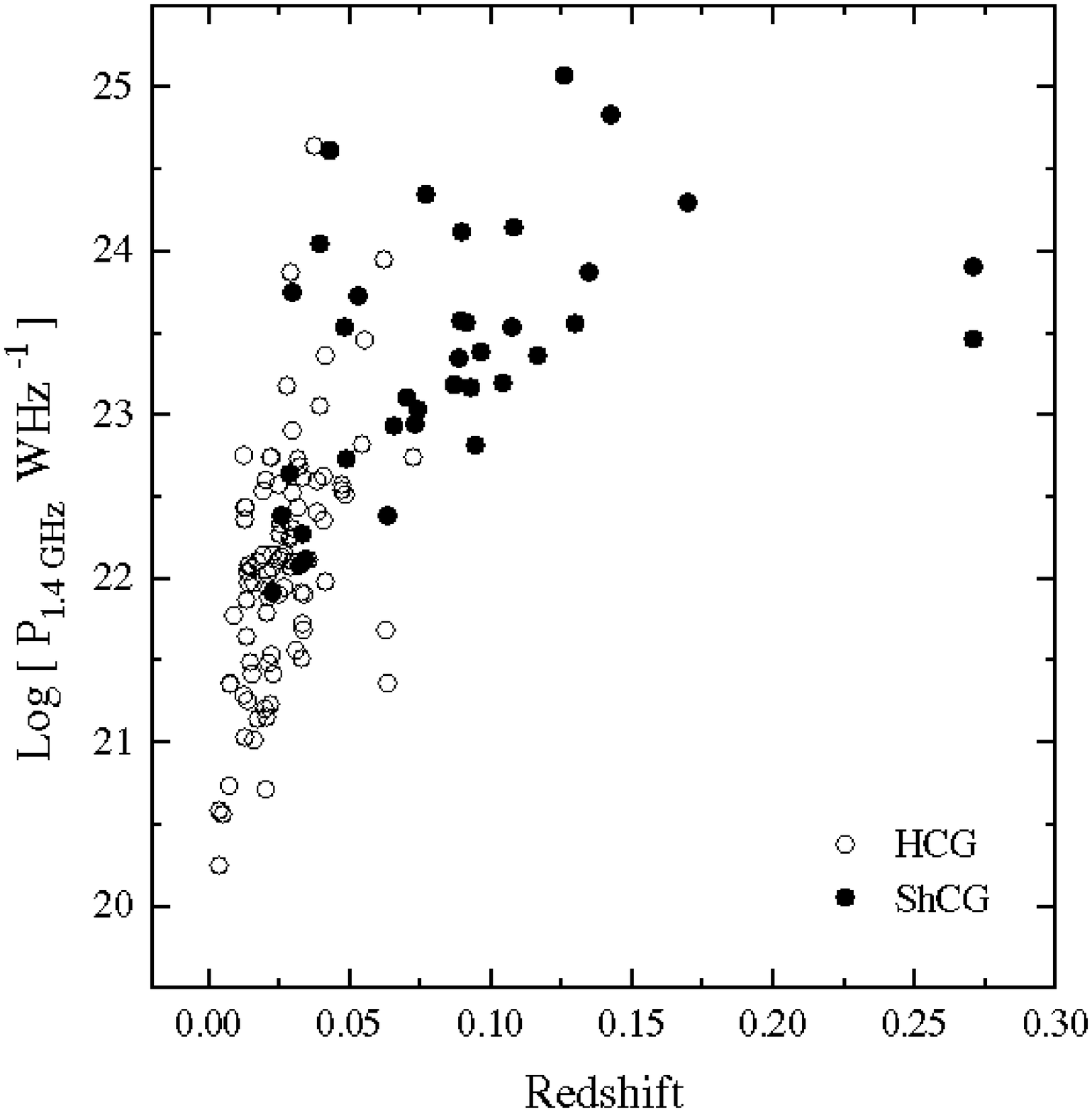}
\end{figure}

\begin{figure}[htb]
\setcounter{figure}{3}
\caption{The comparison of the radio and 60 $\mu$m fluxes of member galaxies in HCGs and ShCGs (the log$F_{60}$-log$F_{1.4}$ graph).}
\epsfxsize=15cm\epsfbox{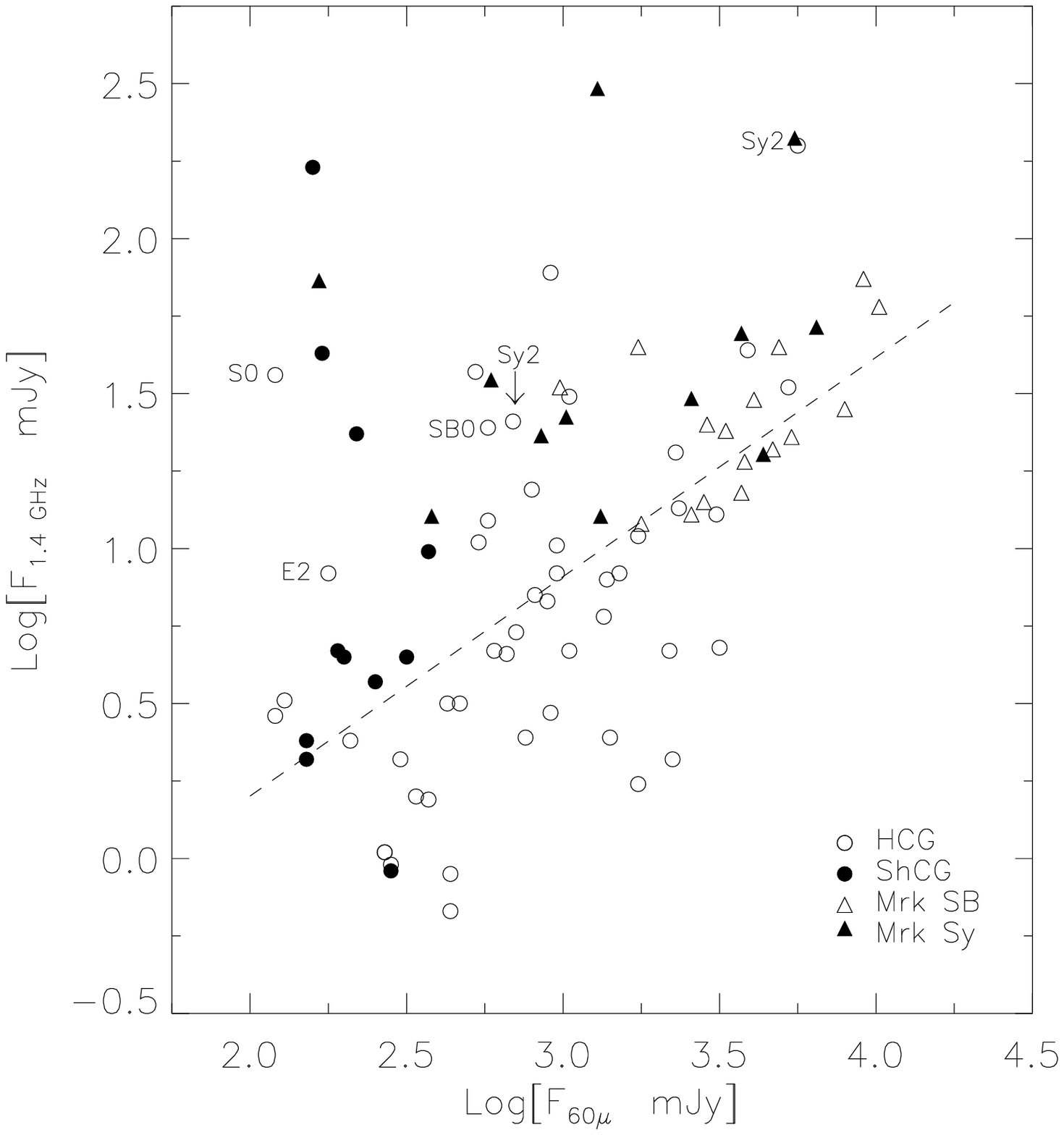}
\end{figure}

\end{document}